\documentclass{elsart}
\usepackage{graphicx}
\begin{document}
\begin{frontmatter}
\title{Searching for a New Source(s) of T-Violation in Spin
Dependent Total Cross Section Measurements}

\author{Guanghua Xu\corauthref{cor1}}
\ead{gxu@uh.edu}
\corauth[cor1]{Corresponding Author}
\address{Department of Physics, 
        University of Houston,
        4800 Calhoun Rd,
        Houston,  TX  USA 77204}
\begin{abstract}
%

We first re-prove with a more complete method that the the minimum standard
model, with the inclusion of
the CKM-matrix, requires the T-odd/P-odd total cross section of two spin-1/2
particles to vanish in all orders\cite{gxu}.  Then
we study the contribution to T-odd/P-odd total scattering cross sections
from various channels within the Higgs sector, and optimize conditions for
possible experimental measurements of these effects. These studies
show that such contributions can appear at tree level, and that the
spin dependent cross section asymmetry is measurable if the lightest
Higgs paarticle is not too massive, {\it e.g.} $m_H \sim 200\, GeV$, and if
suitable reaction channels and beam energies and luminosities are chosen.
\end{abstract}

\end{frontmatter}
\setcounter{equation}{0}
\section{Introduction}
 
The minimal standard model\cite{weinbg} with the
Cabibbo-Kobayashi-Maskawa
mixing matrix\cite{km}, MSMCKM, explains CP-violations in heavy
quark decays. However, it is natural to wonder if other
CP-violations are possible. Various attempts to explain
the baryon asymmetry of the universe require much larger
CP-violation\cite{dolgov} than suggested by the MSMCKM,
which may indicate additional CP or T
violating mechanisms. Indeed, the possibility of CP (or T) violation due to the
Higgs sector has been independently studied by the authors in refs. \cite{lee}
and \cite{weinberg}, and models proposed that introduce CP-violation through
both neutral and charged Higgs-boson exchange\cite{weinberg2}.

Recently, it was shown\cite{gxu} that while the MSMCKM gives a null result
to all orders for an experimental test of time-reversal symmetry
($1/2 + 1/2 \rightarrow 1/2 + 1/2$) as suggested by ref. \cite{conzett},
this is not necessarily true if the Higgs sector
also contributes to T (or CP) violation.
Thus, ref. \cite{gxu} indicates that such a measurement is a
null test of extensions to the MSMCKM which could include
CP (or T)-violation contributed by the Higgs sector. However,
from an experimental view point, one must know whether a test is
feasible and what precision would be required. In what follows, we
study several possible experimental tests. We propose
1) appropriate
reaction channels for an experimental measurement, and 2)
appropriate reaction energies.

In section 2, we re-prove the theorem that the
T-odd/P-odd total cross section in the MSMCKM vanishes to all orders
\cite{gxu} by a different, and more complete, method.
In section 3, the T-odd/P-odd total cross sections for various possible
channels in a model extended to include contributions of the Higgs sector
to T (or CP) violation,  are studied as a function of
the beam energy, in order to optimize the conditions for an experimental
test. Finally, conclusions are given in section 4.

\section{Proof of Null T-odd/P-odd $A_{x,y}$ within the MSMCKM}

This theorem was originally proven in ref. \cite{gxu}.  Now consider
the reaction of two spin-1/2 particles
(${1 \over 2} + {1 \over 2} \rightarrow {1 \over 2} + {1 \over 2}$).
The forward-scattering matrix element is written as,\cite{conzett}
\begin{eqnarray}
  M(0) & = & a_{0,0} + a_{0,z} \sigma_0 \sigma_z + a_{z,0} \sigma_z \sigma_0 +
         a_{x,x} (\sigma_x \sigma_x +  \sigma_y \sigma_y ) + \nonumber \\
       &   & a_{z,z} \sigma_z \sigma_z + a_{x,y} (\sigma_x \sigma_y -
         \sigma_y \sigma_x)\, ,
\end{eqnarray}
where $\sigma_x \equiv {\bf \sigma \cdot x}$, etc., and
we have chosen a coordinate system by defining the unit vectors,
\begin{eqnarray}
  {\bf e_z} &  = & {\bf k_z} /k_z \nonumber \\
  {\bf e_y} & = & {\bf k \times k^{\prime}}/|{\bf k \times
  k^{\prime}}| \nonumber \\
  {\bf e_x} & = & {\bf e_y} \times {\bf e_z} .
\end{eqnarray}
Here ${\bf k}$ and ${\bf k^\prime }$ are the incident and scattered
momenta of the particles, respectively.  The conditions for
parity-non-conserving (PNC)
and time-reversal-volated (TRV) amplitudes have the properties:
\begin{eqnarray}
  {\mathrm{PNC\ (TRV)\ if\ }}n_x +n_z\ (n_x ){\mathrm{\ is\ odd}}.
\end{eqnarray}
In this equation $n_x (n_z)$ is the number of $x(z)$ subscripts. Thus, the
only TRV
amplitude in eq. (1) is $a_{xy}$, which is also PNC, {\it i.e.} T-odd/P-odd.

Using the optical theorem, which relates the total cross section to the
imaginary part of the forward-scattering amplitude, the total
spin-correlation coefficient $A_{x,y}$ for $p_1 = p_x$ and $p_2 = \pm p_y$
is given by,
\begin{eqnarray}
  A_{x,y} = Im a_{x,y} / Im a_{0,0}\, .
\end{eqnarray}
Therefore, this spin-correlation coefficient $A_{x,y}$ is both time (T) and
parity (P) odd. A non-zero $A_{x,y}$ indicates not only that the
reaction violates T and P but also that the T-odd/P-odd total cross section
is not zero. This is a null test, and provides a framework to precisely
investigate TRV processes.

The above derivation is obtained from the two-component spinor description,
but the conclusions obtained using eq. (1) to eq. (4) can be applied
in a four-component relativistic description, provided the center-of-mass frame
(CMS) is used. This follows from\cite{stapp}
1) the four-component relativistic scattering matrix can be reduced to the
two-component formalism in the center-of-mass frame (CMS); 2)
the reduced Pauli scattering matrix has the same transformation properties
as the non-relativistic scattering matrix under spatial reflections and time
reversal; and 3) the spin vector in the non-relativistic treatment can be
equated
to the relativistic spin vector when the latter is measured in the particle
rest-frame related to the CMS by a Lorentz transformation.

Ref. \cite{gxu} proved that $A_{x,y}$
of polarized scattering $1/2 + 1/2 \rightarrow 1/2 + 1/2$
in the MSMCKM is identically
zero. For a more complete  proof, we note that in
the MSMCKM, CP(T)-violation is caused by mixing among the
three generations of quarks. A complex phase, $\delta$, in the CKM-matrix
provides a natural mechanism for the small, but non-zero violation, of
CP conservation. The T (or CP) violating components of the Lagrangian are
contained in the expression,
\begin{eqnarray}
  {\mathcal L} = {g \over  \sqrt{2} } ( J_{\mu}^+ W^{+\mu} +
             J^-_{\mu} W^{-\mu} ) \ ,
\end{eqnarray}
where $g$ is the coupling constant, $W^{\pm}_{\mu}$ are the charged vector
bosons, and $J^{\pm}_{\mu}$ are the $SU(2)$ fermionic currents. These
have the values,
\begin{eqnarray}
  g & = & e \sin \theta_W \ , \ \ \ \ \ \ \ \
  W^{\pm}_{\mu} = (A^1_{\mu} \mp i A^2_{\mu} ) / \sqrt{2} \ , \nonumber \\
  J^+_{\mu} & = & J^1_{\mu} + i J^2_{\mu}
                = {1 \over 2} \overline{U} \gamma_{\mu} (1- \gamma^5 )VD \ ,
  \ \ \ \ \ \ \ \ J^-_{\mu} = J^{+\dagger}_{\mu} \ .
\end{eqnarray}
Here $V$ is the CKM mixing matrix, $\theta_W$ is the Weinberg angle, and $U$
and $D$ are quark triplets $(u,\,c,\,t)$ and $(d,\,s,\,b)$, respectively.
Based on eqs. (5) and (6), only the scatterings of quarks can
possibly introduce T (or CP) vioation components.

At tree level, the Feynman Diagrams of the forward scattering amplitude which
could possibly contribute to T-odd total cross section are,

\begin{picture}(333,147)
  \multiput(30,127)(45,0){2}{\vector(1,0){22.5}}
  \multiput(52.5,127)(45,0){2}{\line(1,0){22.5}}
  \multiput(75,124.5)(0,-10){5}{\oval(5,5)[r]}
  \multiput(75,119.5)(0,-10){5}{\oval(5,5)[l]}
  \multiput(30,77)(45,0){2}{\vector(1,0){22.5}}
  \multiput(52.5,77)(45,0){2}{\line(1,0){22.5}}
  \put(83,92){$W^+$}
  \put(75,127){\circle*{3}}
  \put(75,77){\circle*{3}}
  \put(72.5,98.5){\vector(0,1){3}}
  \put(27,132){$a$}
  \put(115,132){$b$}
  \put(27,62){$b$}
  \put(115,62){$a$}
  \put(70,135){$V_{ab}$}
  \put(70,62){$V_{ab}^*$}
  \put(75,44){(i)}
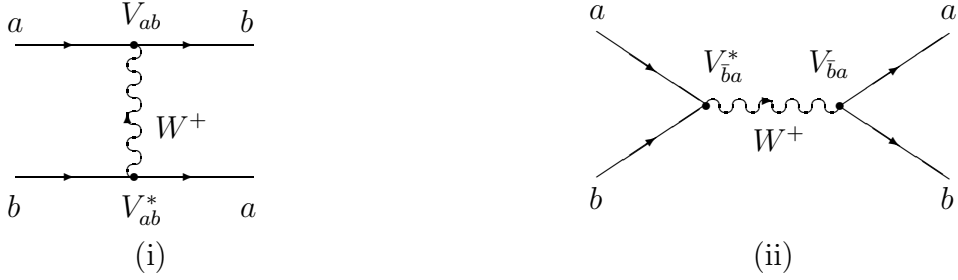
  \put(0,22){Fig. 1: The forward scattering amplitudes
  ${\mathcal M}(ab\longrightarrow ab)$ at tree level that}
  \put(40,0){could possibly contribute to T-odd total cross section.}
  \put(250,132){\vector(3,-2){22.5}}
  \put(268.72,119.52){\line(3,-2){22.5}}
  \multiput(293.92,104)(10,0){5}{\oval(5,5)[t]}
  \multiput(298.92,104)(10,0){5}{\oval(5,5)[b]}
  \put(250,77){\vector(3,2){22.5}}
  \put(268.72,89.48){\line(3,2){22.5}}
  \put(342.,104){\vector(3,2){22.5}}
  \put(342.,104){\vector(3,-2){22.5}}
  \put(360.71,116.5){\line(3,2){22.5}}
  \put(360.71,91.5){\line(3,-2){22.5}}
  \put(309,87){$W^+$}
  \put(291.42,104){\circle*{3}}
  \put(342,104){\circle*{3}}
  \put(313.5,105.9){\vector(1,0){3}}
  \put(247,137){$a$}
  \put(380,137){$a$}
  \put(247,65){$b$}
  \put(380,65){$b$}
  \put(290,117){$V_{\bar{b}a}^*$}
  \put(330,117){$V_{\bar{b}a}$}
  \put(309,44){(ii)}
\end{picture}

The forward scattering amplitude for Fig. 1(i) (${\mathcal M}_{1i}$) and
Fig. 1(ii) (${\mathcal M}_{1ii}$) in Feynman-'t Hooft gauge are given by,
\begin{eqnarray}
  {\mathcal M}_{1i} & = & {-g^2 \over 8}\, {1\over k_i^2 -M_W^2 +i \epsilon}
    V_{ab} V_{ab}^{*} \cdot
    [ \overline{u}_b \, \gamma^{\mu} (1-\gamma^5 ) \, u_a ]
[ \overline{u}_a \, \gamma_{\mu} (1-\gamma^5 ) \, u_b ] \, ; \nonumber \\
  {\mathcal M}_{1ii} & = & {-g^2 \over 8}\, {1\over k_{ii}^2 -M_W^2 +i\epsilon}
    V_{\bar{b}a}^* V_{\bar{b}a} \cdot
   [ \overline{u}_{\bar{b}} \, \gamma^{\mu} (1-\gamma^5 )
  \, u_a ] [ \overline{u}_a \, \gamma_{\mu} (1-\gamma^5 ) \, u_{\bar{b}} ] \, ;
\end{eqnarray}
with $k_i^2 = (p_a -p_b)^2$ and $k_{ii}^2 = (p_a +p_b)^2$.

From the T-even condition, ${\mathcal M}^T = {\mathcal M}^{\dagger}$,
one can obtain the T-even and T-odd amplitudes as,
\begin{eqnarray}
  &   & {\mathcal M}_{1i}^{even} = {-g^2 \over 8}\, {1\over k_a^2 -M_W^2 +
    i\epsilon} Re(V_{ab} V_{ab}^{*})
    [ \overline{u}_b \, \gamma^{\mu} (1-\gamma^5 ) \, u_a ]
    [ \overline{u}_a \, \gamma_{\mu} (1-\gamma^5 ) \, u_b ] \, ; \nonumber \\
  &   & {\mathcal M}_{1ii}^{even} = {-g^2 \over 8}\, {1\over k_b^2 -M_W^2 +
    i\epsilon} Re(V_{\bar{b}a}^* V_{\bar{b}a})
    [ \overline{u}_{\bar{b}} \, \gamma^{\mu} (1-\gamma^5 ) \, u_a ]
    [ \overline{u}_a \, \gamma_{\mu} (1-\gamma^5 ) \, u_{\bar{b}} ] \, ;
  \nonumber \\
  &   & {\mathcal M}_{1i}^{odd} = {-g^2 \over 8}\, {1\over k_a^2 -M_W^2 +
    i\epsilon} i Im(V_{ab} V_{ab}^{*})
    [ \overline{u}_b \, \gamma^{\mu} (1-\gamma^5 ) \, u_a ]
    [ \overline{u}_a \, \gamma_{\mu} (1-\gamma^5 ) \, u_b ] \, ;\nonumber \\
  &   & {\mathcal M}_{1ii}^{odd} = {-g^2 \over 8}\, {1\over k_b^2 -M_W^2 +
    i\epsilon} i Im(V_{\bar{b}a}^* V_{\bar{b}a})
    [ \overline{u}_{\bar{b}} \, \gamma^{\mu} (1-\gamma^5 )
    \, u_a ] [ \overline{u}_a \, \gamma_{\mu} (1-\gamma^5 ) \, u_{\bar{b}} ]
    \, .\nonumber \\
\end{eqnarray}

Since $Im(V_{ab} V_{ab}^{*}) = 0$ and $Im (V_{\bar{b}a}^* V_{\bar{b}a}) = 0$,
the T-odd amplitudes are zero at tree level.

At one-loop level, the Feynman diagrams of the forward scattering amplitudes
which
could possibly contribute to T-odd total cross section are the following.

\begin{picture}(333,250)
  \multiput(5,235)(64,0){2}{\vector(1,0){12.5}}
  \multiput(17.5,235)(64,0){2}{\line(1,0){12.5}}
  \put(30,235){\vector(1,0){19.5}}
  \put(49.5,235){\line(1,0){19.5}}
  \multiput(30,227.5)(0,-10){5}{\oval(5,5)[r]}
  \multiput(30,232.5)(0,-10){6}{\oval(5,5)[l]}
  \multiput(69,227.5)(0,-10){5}{\oval(5,5)[r]}
  \multiput(69,232.5)(0,-10){6}{\oval(5,5)[l]}
  \multiput(5,180)(64,0){2}{\vector(1,0){12.5}}
  \multiput(17.5,180)(64,0){2}{\line(1,0){12.5}}
  \put(30,180){\vector(1,0){19.5}}
  \put(49.5,180){\line(1,0){19.5}}
  \put(5,205){$W^+$}
  \put(73,205){$W^+$}
  \put(30,235){\circle*{3}}
  \put(30,180){\circle*{3}}
  \put(69,235){\circle*{3}}
  \put(69,180){\circle*{3}}
  \put(32.5,206.5){\vector(0,1){3}}
  \put(72,208.5){\vector(0,-1){3}}
  \put(2,240){$a$}
  \put(90,240){$a$}
  \put(2,165){$b$}
  \put(90,165){$b$}
  \put(48,240){$x$}
  \put(48,165){$y$}
  \put(25,243){$V_{ax}$}
  \put(67,243){$V_{ax}^*$}
  \put(25,165){$V_{yb}^*$}
  \put(67,165){$V_{yb}$}
  \put(48,145){(i)}
  \put(135,235){\vector(1,0){12.5}}
  \put(147.5,235){\line(1,0){12.5}}
  \put(199,235){\vector(1,-1){50}}
  \put(160,235){\vector(1,0){19.5}}
  \put(179.5,235){\line(1,0){19.5}}
  \multiput(160,227.5)(0,-10){5}{\oval(5,5)[r]}
  \multiput(160,232.5)(0,-10){6}{\oval(5,5)[l]}
  \multiput(199,227.5)(0,-10){5}{\oval(5,5)[r]}
  \multiput(199,232.5)(0,-10){6}{\oval(5,5)[l]}
  \put(135,180){\vector(1,0){12.5}}
  \put(147.5,180){\line(1,0){12.5}}
  \put(199,180){\vector(1,1){50}}
  \put(160,180){\vector(1,0){19.5}}
  \put(179.5,180){\line(1,0){19.5}}
  \put(135,205){$W^+$}
  \put(171,205){$Z(\gamma)$}
  \put(160,235){\circle*{3}}
  \put(160,180){\circle*{3}}
  \put(199,235){\circle*{3}}
  \put(199,180){\circle*{3}}
  \put(162.5,206.5){\vector(0,1){3}}
  \put(202,208.5){\vector(0,-1){3}}
  \put(132,240){$a$}
  \put(240,175){$b$}
  \put(132,165){$b$}
  \put(240,230){$a$}
  \put(178,240){$b$}
  \put(178,165){$a$}
  \put(155,243){$V_{ab}$}
  \put(155,165){$V_{ab}^*$}
  \put(178,145){(ii)}
  \put(280,235){\vector(1,0){12.5}}
  \put(292.5,235){\line(1,0){12.5}}
  \put(344,235){\vector(1,-1){50}}
  \put(305,235){\vector(1,0){19.5}}
  \put(324.5,235){\line(1,0){19.5}}
  \multiput(305,227.5)(0,-10){5}{\oval(5,5)[r]}
  \multiput(305,232.5)(0,-10){6}{\oval(5,5)[l]}
  \multiput(344,227.5)(0,-10){5}{\oval(5,5)[r]}
  \multiput(344,232.5)(0,-10){6}{\oval(5,5)[l]}
  \put(280,180){\vector(1,0){12.5}}
  \put(292.5,180){\line(1,0){12.5}}
  \put(344,180){\vector(1,1){50}}
  \put(305,180){\vector(1,0){19.5}}
  \put(324.5,180){\line(1,0){19.5}}
  \put(275,205){$Z(\gamma)$}
  \put(319,205){$W^+$}
  \put(305,235){\circle*{3}}
  \put(305,180){\circle*{3}}
  \put(344,235){\circle*{3}}
  \put(344,180){\circle*{3}}
  \put(307.5,208.5){\vector(0,-1){3}}
  \put(347,206.5){\vector(0,1){3}}
  \put(277,240){$a$}
  \put(385,230){$a$}
  \put(277,165){$b$}
  \put(323,165){$b$}
  \put(323,240){$a$}
  \put(385,175){$b$}
  \put(342,243){$V_{ab}$}
  \put(342,165){$V_{ab}^*$}
  \put(323,145){(iii)}
  \put(100,120){\vector(3,-2){22.5}}
  \put(118.72,107.52){\line(3,-2){22.5}}
  \multiput(143.92,92)(10,0){5}{\oval(5,5)[t]}
  \multiput(148.92,92)(10,0){5}{\oval(5,5)[b]}
  \put(100,65){\vector(3,2){22.5}}
  \put(118.72,77.48){\line(3,2){22.5}}
  \put(192.,92){\vector(3,2){22.5}}
  \put(192.,92){\vector(3,-2){22.5}}
  \put(210.71,104.5){\line(3,2){22.5}}
  \put(210.71,79.5){\line(3,-2){22.5}}
  \put(159,75){$W^+$}
  \put(141.42,92){\circle*{3}}
  \put(192,92){\circle*{3}}
  \put(163.5,93.9){\vector(1,0){3}}
  \put(97,125){$a$}
  \put(230,125){$a$}
  \put(97,53){$b$}
  \put(230,53){$b$}
  \put(130,105){$V_{\bar{b}a}^*$}
  \put(180,105){$V_{\bar{b}a}$}
  \put(159,45){(iv)}
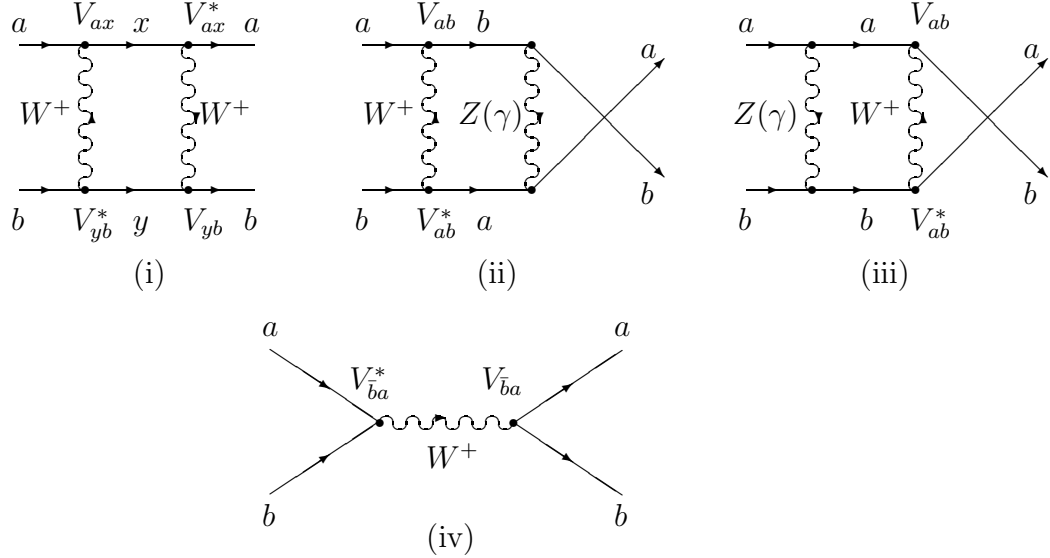
  \put(0,22){Fig. 2: The forward scattering amplitudes
  ${\mathcal M}(ab\longrightarrow ab)$ at one-loop level that}
  \put(39,3){could possibly contribute to T-odd total cross section.}
\end{picture}

Since
\begin{eqnarray}
  [\bar{u}_c \gamma^\mu (1-\gamma^5) u_a
    \bar{u}_d \gamma^\mu (1-\gamma^5) u_b]^T
  = [\bar{u}_c \gamma^\mu (1-\gamma^5) u_a
    \bar{u}_d \gamma^\mu (1-\gamma^5) u_b]^\dagger \ ,
\end{eqnarray}
the only factor that determines T-odd or T-even is the CKM-matrix
elements in the amplitudes. The possible T-odd factors in
Fig. 2(ii, iii) are the same as the factors in Fig. 1 and give zero T-odd
amplitudes, {\it i.e.} the vertices that do not include CKM-matrix elements
do not introduce T-odd factors. For Fig. 2(i), the multiplication of the
matrix elements is given by,
\begin{eqnarray}
  V_{xa} V_{xa}^* V_{by}^* V_{by} = |V_{xa} V_{by}|^2 = {\mathrm{real\ and\
  therefore}}\ Im(V_{xa} V_{xa}^* V_{by}^* V_{by}) = 0\, .
\end{eqnarray}
For Fig. 2(iv), the multiplication of the matrix elements is given by
\begin{eqnarray}
  V_{\bar{b}a} V_{\bar{y}x}^* V_{\bar{y}x} V_{\bar{b}a}^* =
    |V_{\bar{b}a} V_{\bar{y}x}|^2 = {\mathrm{real\ and\ therefore}} \
    Im(V_{\bar{b}a} V_{\bar{y}x}^* V_{\bar{y}x} V_{\bar{b}a}^* ) = 0\, .
\end{eqnarray}

Therefore, at one-loop level, T-odd amplitude is zero.

For arbitrary $n$-th order, the forward scattering $a+b \rightarrow a+b$
would go through combinations of the following processes as shown in
Figs. 3 and Fig. 4, depending on the particles $a$ and $b$.
\cite{expl1}

\begin{picture}(433,250)
  \multiput(5,235)(25,0){3}{\vector(1,0){12.5}}
  \multiput(17.5,235)(25,0){3}{\line(1,0){12.5}}
  \multiput(83,232.5)(5,0){3}{$\cdot$}
  \multiput(100,235)(25,0){2}{\line(1,0){12.5}}
  \put(112.5,235){\vector(1,0){12.5}}
  \put(135,235){\vector(1,-1){50}}
  \multiput(28,227.5)(0,-10){2}{\oval(5,5)[r]}
  \multiput(28,232.5)(0,-10){2}{\oval(5,5)[l]}
  \multiput(53,227.5)(0,-10){2}{\oval(5,5)[r]}
  \multiput(53,232.5)(0,-10){2}{\oval(5,5)[l]}
  \multiput(78,227.5)(0,-10){2}{\oval(5,5)[r]}
  \multiput(78,232.5)(0,-10){2}{\oval(5,5)[l]}
  \multiput(110,227.5)(0,-10){2}{\oval(5,5)[r]}
  \multiput(110,232.5)(0,-10){2}{\oval(5,5)[l]}
  \multiput(135,227.5)(0,-10){2}{\oval(5,5)[r]}
  \multiput(135,232.5)(0,-10){2}{\oval(5,5)[l]}
  \multiput(27,235)(25,0){3}{\circle*{3}}
  \multiput(110,235)(25,0){2}{\circle*{3}}
  \multiput(27,198)(0,5){3}{$\cdot$}
  \multiput(52,198)(0,5){3}{$\cdot$}
  \multiput(77,198)(0,5){3}{$\cdot$}
  \multiput(109,198)(0,5){3}{$\cdot$}
  \multiput(134,198)(0,5){3}{$\cdot$}
  \multiput(28,190)(0,-10){2}{\oval(5,5)[l]}
  \multiput(28,195)(0,-10){2}{\oval(5,5)[r]}
  \multiput(53,190)(0,-10){2}{\oval(5,5)[l]}
  \multiput(53,195)(0,-10){2}{\oval(5,5)[r]}
  \multiput(78,190)(0,-10){2}{\oval(5,5)[l]}
  \multiput(78,195)(0,-10){2}{\oval(5,5)[r]}
  \multiput(110,190)(0,-10){2}{\oval(5,5)[l]}
  \multiput(110,195)(0,-10){2}{\oval(5,5)[r]}
  \multiput(135,190)(0,-10){2}{\oval(5,5)[l]}
  \multiput(135,195)(0,-10){2}{\oval(5,5)[r]}
  \multiput(5,177.5)(25,0){3}{\vector(1,0){12.5}}
  \multiput(17.5,177.5)(25,0){3}{\line(1,0){12.5}}
  \multiput(83,175)(5,0){3}{$\cdot$}
  \multiput(100,177.5)(25,0){2}{\line(1,0){12.5}}
  \put(112.5,177.5){\vector(1,0){12.5}}
  \put(135,177.5){\vector(1,1){50}}
  \multiput(27,177.5)(25,0){3}{\circle*{3}}
  \multiput(110,177.5)(25,0){2}{\circle*{3}}
  \put(0,225){$a$}
  \put(0,180){$b$}
  \put(35,225){$x_1$}
  \put(35,184){$y_1$}
  \put(60,225){$x_2$}
  \put(60,184){$y_2$}
  \put(115,225){$ x_n$}
  \put(115,184){$y_n$}
  \multiput(30.5,193)(50,0){2}{\vector(0,1){3}}
  \put(55.5,195){\vector(0,-1){3}}
  \multiput(30.5,217)(50,0){2}{\vector(0,1){3}}
  \put(55.5,218){\vector(0,-1){3}}
  \put(112.5,195){\vector(0,-1){3}}
  \put(137.5,194){\vector(0,1){3}}
  \put(112.5,218){\vector(0,-1){3}}
  \put(137.5,217){\vector(0,1){3}}
  \put(185,230){$a$}
  \put(185,175){$b$}
  \put(18,243){$V_{ax_1}$}
  \put(18,162.5){$V_{y_1b}^*$}
  \put(45,243){$V_{x_2 x_1}^*$}
  \put(45,162.5){$V_{y_1 y_2}$}
  \put(72,243){$V_{x_2 x_3}$}
  \put(72,162.5){$V_{y_3 y_2}^*$}
  \put(99,243){$V_{x_n x_{n-1}}^*$}
  \put(99,162.5){$V_{y_{n-1} y_n}$}
  \put(137,243){$V_{x_n b}$}
  \put(137,162.5){$V_{a y_n}^*$}
  \multiput(7,205)(25,0){3}{$W^+$}
  \multiput(90,205)(25,0){2}{$W^+$}
  \put(80,140){(i)}
  \multiput(235,235)(25,0){3}{\vector(1,0){12.5}}
  \multiput(247.5,235)(25,0){3}{\line(1,0){12.5}}
  \multiput(313,232.5)(5,0){3}{$\cdot$}
  \multiput(330,235)(25,0){2}{\line(1,0){12.5}}
  \multiput(342.5,235)(25,0){2}{\vector(1,0){12.5}}
  \multiput(258,227.5)(0,-10){2}{\oval(5,5)[r]}
  \multiput(258,232.5)(0,-10){2}{\oval(5,5)[l]}
  \multiput(283,227.5)(0,-10){2}{\oval(5,5)[r]}
  \multiput(283,232.5)(0,-10){2}{\oval(5,5)[l]}
  \multiput(308,227.5)(0,-10){2}{\oval(5,5)[r]}
  \multiput(308,232.5)(0,-10){2}{\oval(5,5)[l]}
  \multiput(340,227.5)(0,-10){2}{\oval(5,5)[r]}
  \multiput(340,232.5)(0,-10){2}{\oval(5,5)[l]}
  \multiput(365,227.5)(0,-10){2}{\oval(5,5)[r]}
  \multiput(365,232.5)(0,-10){2}{\oval(5,5)[l]}
  \multiput(257,235)(25,0){3}{\circle*{3}}
  \multiput(340,235)(25,0){2}{\circle*{3}}
  \multiput(257,198)(0,5){3}{$\cdot$}
  \multiput(282,198)(0,5){3}{$\cdot$}
  \multiput(307,198)(0,5){3}{$\cdot$}
  \multiput(339,198)(0,5){3}{$\cdot$}
  \multiput(364,198)(0,5){3}{$\cdot$}
  \multiput(258,190)(0,-10){2}{\oval(5,5)[l]}
  \multiput(258,195)(0,-10){2}{\oval(5,5)[r]}
  \multiput(283,190)(0,-10){2}{\oval(5,5)[l]}
  \multiput(283,195)(0,-10){2}{\oval(5,5)[r]}
  \multiput(308,190)(0,-10){2}{\oval(5,5)[l]}
  \multiput(308,195)(0,-10){2}{\oval(5,5)[r]}
  \multiput(340,190)(0,-10){2}{\oval(5,5)[l]}
  \multiput(340,195)(0,-10){2}{\oval(5,5)[r]}
  \multiput(365,190)(0,-10){2}{\oval(5,5)[l]}
  \multiput(365,195)(0,-10){2}{\oval(5,5)[r]}
  \multiput(235,177.5)(25,0){3}{\vector(1,0){12.5}}
  \multiput(247.5,177.5)(25,0){3}{\line(1,0){12.5}}
  \multiput(313,175)(5,0){3}{$\cdot$}
  \multiput(330,177.5)(25,0){2}{\line(1,0){12.5}}
  \multiput(342.5,177.5)(25,0){2}{\vector(1,0){12.5}}
  \multiput(257,177.5)(25,0){3}{\circle*{3}}
  \multiput(340,177.5)(25,0){2}{\circle*{3}}
  \put(230,225){$a$}
  \put(230,180){$b$}
  \put(265,225){$x_1$}
  \put(265,184){$y_1$}
  \put(290,225){$x_2$}
  \put(290,184){$y_2$}
  \put(345,225){$ x_n$}
  \put(345,184){$y_n$}
  \put(380,225){$a$}
  \put(380,180){$b$}
  \multiput(260.5,193)(50,0){2}{\vector(0,1){3}}
  \put(285.5,195){\vector(0,-1){3}}
  \multiput(260.5,217)(50,0){2}{\vector(0,1){3}}
  \put(285.5,218){\vector(0,-1){3}}
  \put(342.5,194){\vector(0,1){3}}
  \put(367.5,195){\vector(0,-1){3}}
  \put(342.5,217){\vector(0,1){3}}
  \put(367.5,218){\vector(0,-1){3}}
  \put(248,243){$V_{ax_1}$}
  \put(248,162.5){$V_{y_1b}^*$}
  \put(275,243){$V_{x_2 x_1}^*$}
  \put(275,162.5){$V_{y_1 y_2}$}
  \put(302,243){$V_{x_2 x_3}$}
  \put(302,162.5){$V_{y_3 y_2}^*$}
  \put(329,243){$V_{x_{n-1} x_n}$}
  \put(329,162.5){$V_{y_n y_{n-1}}^*$}
  \put(367,243){$V_{a x_n}^*$}
  \put(367,162.5){$V_{y_n b}$}
  \multiput(237,205)(25,0){3}{$W^+$}
  \multiput(320,205)(25,0){2}{$W^+$}
  \put(310,140){(ii)}
  \put(100,120){\vector(3,-2){22.5}}
  \put(118.72,107.52){\line(3,-2){22.5}}
  \multiput(143.92,92)(10,0){3}{\oval(5,5)[t]}
  \multiput(148.92,92)(10,0){3}{\oval(5,5)[b]}
  \put(100,65){\vector(3,2){22.5}}
  \put(118.72,77.48){\line(3,2){22.5}}
  \put(183.92,92){\circle{25}}
  \multiput(198.92,92)(10,0){3}{\oval(5,5)[t]}
  \multiput(203.92,92)(10,0){3}{\oval(5,5)[b]}
  \put(226.42,92){\vector(3,2){22.5}}
  \put(226.42,92){\vector(3,-2){22.5}}
  \put(245.14,104.5){\line(3,2){22.5}}
  \put(245.14,79.5){\line(3,-2){22.5}}
  \put(150,75){$W^+$}
  \put(202.5,75){$W^+$}
  \put(141.42,92){\circle*{3}}
  \put(226,92){\circle*{3}}
  \put(153.5,93.9){\vector(1,0){3}}
  \put(208.5,93.9){\vector(1,0){3}}
  \put(97,125){$a$}
  \put(264.42,125){$a$}
  \put(97,53){$b$}
  \put(264.42,53){$b$}
  \put(130,105){$V_{\bar{b}a}^*$}
  \put(214,105){$V_{\bar{b}a}$}
  \multiput(177,99)(5,0){3}{$\cdot$}
  \multiput(174,94)(5,0){4}{$\cdot$}
  \multiput(173,89)(5,0){5}{$\cdot$}
  \multiput(173,84)(5,0){5}{$\cdot$}
  \multiput(181,79)(5,0){2}{$\cdot$}
  \put(159,45){(iii)}
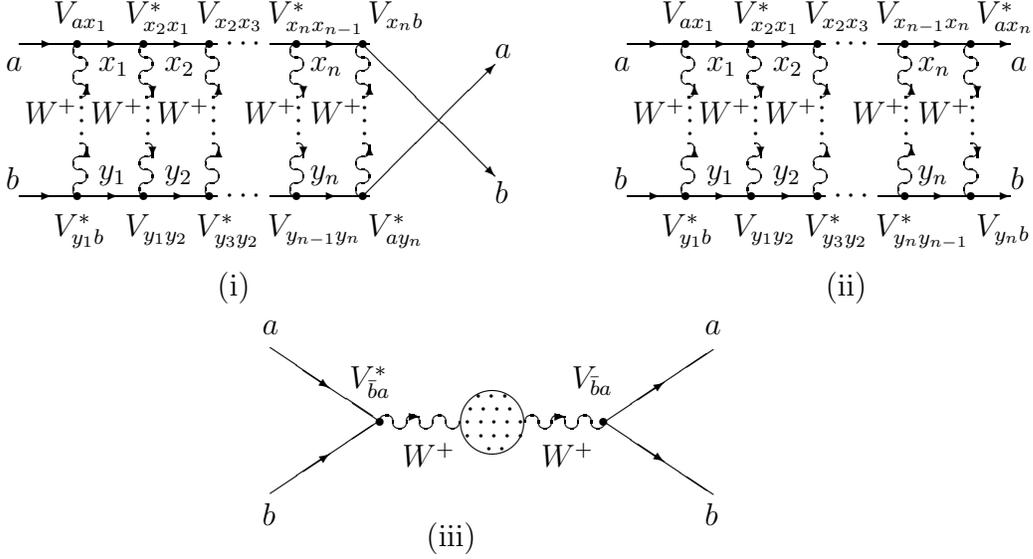
  \put(0,22){Fig. 3: The forward elastic scattering amplitudes
  ${\mathcal M}(ab\longrightarrow ab)$ at $n$-th order}
  \put(39,3){that could possibly contribute to T-odd total cross section.}
\end{picture}

\begin{picture}(300,60)
  \multiput(5,10)(50,0){3}{\vector(1,0){25}}
  \multiput(30,10)(50,0){3}{\line(1,0){25}}
  \multiput(55,10)(5,5){5}{\oval(5,5)[tl]}
  \multiput(55,15)(5,5){5}{\oval(5,5)[br]}
  \put(80,34){\oval(5,5)[t]}
  \multiput(85,35)(5,-5){5}{\oval(5,5)[bl]}
  \multiput(85,30)(5,-5){5}{\oval(5,5)[tr]}
  \put(79.5,36.5){\vector(1,0){3}}
  \put(5,0){$a$}
  \put(80,45){$W^+$}
  \put(145,0){$a$}
  \put(80,0){$x$}
  \multiput(52.5,10)(55,0){2}{\circle*{3}}
  \put(33,18){$V_{xa}^*$}
  \put(109,16){$V_{xa}$}
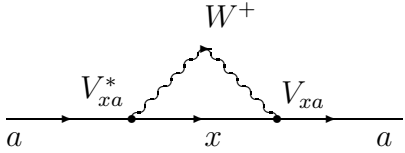
  \put(170,25){Fig. 4: Quark self-mass diagram.}
\end{picture}

If $a$ and $b$ belong to the $U(D)$ and $D(U)$ sectors respectively, the forward
scattering goes through the processes shown in Figs. 3(a) and (b), but
if $a$ and $b$ belong to $U(\bar{D})$ and
$\bar{D}(U)$ sectors respectively or $\bar{U}(D)$ and $D(\bar{U})$ sectors
respectively, the forward scattering goes through the processes shown in
Figs. 3(c).

We prove in the following that the multiplication of CKM matrix
elements in each diagram of Figs. 3 and 4 is real; and since an
arbitrary T-odd amplitude must be a combination of Figs. 3(i, ii, iii) and 4,
the T-odd forward scattering amplitude for an arbitrary order is zero.

Fig. 4 is T-even which is obvious since the CKM-matrix contribution
$V_{xa}V_{xa}^*$ is real.
Without losing generality, let $a$ be in the $D$ sector and $b$ be in the
$U$ sector.
The $n$-th order forward scattering amplitude could go through Figs. 3(i) if
$n$ is an odd number or Figs. 3(ii) if $n$ is an even number.

The amplitude of Fig. 3(i) has the form,
\begin{eqnarray}
  {\mathcal M}_{3i}(0) & \propto & {\mathcal M}_{3i,1}(0) {\mathcal M}_{3i,2}(0)
    \ \ with \nonumber \\
  {\mathcal M}_{3i,1}(0) & \propto &
  (V_{x_n b} \bar{u}_b \gamma^{\mu_{n+1}} (1-\gamma^5) u_{x_n} W^+_{\mu_{n+1}})
  (V_{x_n x_{n-1}}^* \bar{u}_{x_n} \gamma^{\mu_n} (1-\gamma^5) u_{x_{n-1}}
    W^{+\dagger}_{\mu_n} )  \nonumber \\
    &   & (V_{x_{n-2} x_{n-1}} \bar{u}_{x_{n-1}} \gamma^{\mu_{n-1}}
    (1-\gamma^5) u_{x_{n-2}} W^+{\mu_{n-1}} )
    \cdot \cdot \cdot  \nonumber \\
    &   & (V_{x_2 x_3} \bar{u}_{x_3} \gamma^{\mu_3} (1-\gamma^5) u_{x_2}
       W^+_{\mu_3} )
    (V_{x_2 x_1}^* \bar{u}_{x_2} \gamma^{\mu_2} (1-\gamma^5) u_{x_1}
       W^{+\dagger}_{\mu_2} )  \nonumber \\
    &   & (V_{a x_1} \bar{u}_{x_1} \gamma^{\mu_1} (1-\gamma^5) u_a
       W^+_{\mu_1} ) \nonumber \\
  {\mathcal M}_{3i,2} (0) & \propto &
    (V_{a y_n}^* \bar{u}_a \gamma^{\mu_{n+1}} (1-\gamma^5) u_{y_n}
    W^{+\dagger}_{\mu_{n+1}} )
    (V_{y_{n-1} y_n} \bar{u}_{y_n} \gamma^{\mu_n} (1-\gamma^5) u_{y_{n-1}}
    W^+_{\mu_n} )  \nonumber \\
    &   & (V_{y_{n-1} y_{n-2}^*} \bar{u}_{y_{n-1}} \gamma^{\mu_{n-1}}
    (1-\gamma^5) u_{y_{n-2}} W^{+\dagger}_{\mu_{n-1}} )
    \cdot \cdot \cdot  \nonumber \\
    &   & (V_{y_3 y_2}^* \bar{u}_{y_3} \gamma^{\mu_3} (1-\gamma^5) u_{y_2}
      W^{+\dagger}_{\mu_3} )
    (V_{y_1 y_2} \bar{u}_{y_2} \gamma^{\mu_2} (1-\gamma^5) u_{y_1}
      W^+_{\mu_2} )  \nonumber \\
    &   & (V_{y_1 b}^* \bar{u}_{y_1} \gamma^{\mu_1} (1-\gamma^5) u_b
     W^{+\dagger}_{\mu_1} ) \, ,
\end{eqnarray}
where the repeated indices $x_i$ and $y_i$ should be summed over the
particles in the corresponding quark sectors and the corresponding
momenta of the particles should be integrated based on conservation of momenta.

Based on the MSMCKM Lagrangian, $x_i$ and $y_{n-i+1}$ should be in the
same quark sector.\cite{expl1} It is obvious that
\begin{eqnarray}
  {\mathcal M}_{3i,1}(0)={\mathcal M}_{3i,2}^\dagger(0)={\mathcal M}_{3i,2}^*
    (0)\ \ {\mathrm{and}}\ \ {\mathcal M}_{3i,1}(0) {\mathcal M}_{3i,2}(0) =
    {\mathrm{real}}\, .
\end{eqnarray}
Therefore, the T-odd amplitude from Fig. 3(i) is zero.

The amplitude of Fig. 3(ii) has the form,
\begin{eqnarray}
  {\mathcal M}_{3ii}(0) & \propto & {\mathcal M}_{3ii,1}(0)
    {\mathcal M}_{3ii,2}(0)\ \ {\mathrm{with}}\nonumber \\
  {\mathcal M}_{3ii,1}(0) & \propto &
    (V_{a x_n}^* \bar{u}_a \gamma^{\mu_{n+1}} (1-\gamma^5) u_{x_n}
    W^{+\dagger}_{\mu_{n+1}} )
    (V_{x_{n-1} x_n} \bar{u}_{x_n} \gamma^{\mu_n} (1-\gamma^5) u_{x_{n-1}}
    W^+_{\mu_n} ) \nonumber \\
    &   & (V_{x_{n-1} x_{n-2}}^* \bar{u}_{x_{n-1}} \gamma^{\mu_{n-1}}
    (1-\gamma^5) u_{x_{n-2}}
    W^{+\dagger}_{\mu_{n-1}} )
    \cdot \cdot \cdot  \nonumber \\
    &   & (V_{x_2 x_3} \bar{u}_{x_3} \gamma^{\mu_3} (1-\gamma^5) u_{x_2}
     W^+_{\mu_3} )
    (V_{x_2 x_1}^* \bar{u}_2 \gamma^{\mu_2} (1-\gamma^5) u_{x_1}
    W^{+\dagger}_{\mu_2} ) \nonumber \\
    &   & (V_{a x_1} \bar{u}_1 \gamma^{\mu_1} (1-\gamma^5) u_a
    W^+_{\mu_1} ) \nonumber \\
  {\mathcal M}_{3ii,2}(0) & \propto &
    (V_{y_n b} \bar{u}_b \gamma^{\mu_{n+1}} (1-\gamma^5) u_{y_n}
     W^+_{\mu_{n+1}} )
    (V_{y_n y_{n-1}}^* \bar{u}_{y_n} \gamma^{\mu_n} (1-\gamma^5) u_{y_{n-1}}
    W^{+\dagger}_{\mu_n} ) \nonumber \\
    &   & (V_{y_{n-2} y_{n-1}} \bar{u}_{y_{n-1}} \gamma^{\mu_{n-1}}
    (1-\gamma^5) u_{y_{n-2}}
    W^+_{\mu_{n-1}} )
    \cdot \cdot \cdot  \nonumber \\
    &   & (V_{y_3 y_2}^* \bar{u}_{y_3} \gamma^{\mu_3} (1-\gamma^5) u_{y_2}
     W^{+\dagger}_{\mu_3} )
    (V_{y_1 y_2} \bar{u}_{y_2} \gamma^{\mu_2} (1-\gamma^5) u_{y_1}
     W^+_{\mu_2} ) \nonumber \\
    &   & (V_{y_1 b}^* \bar{u}_{y_1} \gamma^{\mu_1} (1-\gamma^5) u_b
     W^{+\dagger}_{\mu_1} ) \, ,
\end{eqnarray}
where the repeated indices $x_i$ and $y_i$ should be summed over the
particles in the corresponding quark sectors and the corresponding
momenta of the particles should be integrated based on conservation of momenta.

Also, based on the MSMCKM Lagrangian, $x_i(y_i) $ and $x_{n-i+1} (y_{n-i+1})$
should be in the same quark sector\cite{expl1} and one should have,
\begin{eqnarray}
  {\mathcal M}_{3ii,l}(0) = {\mathcal M}_{3ii,l}^\dagger (0) =
    {\mathcal M}_{3ii,l}^* (0)\ = {\mathrm{real}}, \ \ l = 1, 2 \, .
\end{eqnarray}
Therefore, the T-odd amplitude from Fig. 3(ii) is zero.

In Fig. 3(iii), a $W^+$ propagator placed after the annihilation of incoming
quarks $a$ and $b$
could create a quark pair in $U$ ($D$) and $\bar{D}$ ($\bar{U}$)
sectors, introducing a possible T-odd contribution. The created $U$ ($D$)
and $\bar{D}$ ($\bar{U}$) pair
will go through the processes of either Fig. 3(i) or Fig. 3(ii) before
the final quark pair in the process is annihilated. This again creates a $W^+$
propagator. \cite{expl2}
Therefore, Fig. 3(iii) can be broken down to a combination
of Fig. 3(i) or Fig. 3(ii), a smaller part of the form of Fig. 3(iii), and
Fig. 4. The smaller part
of Fig. 3(iii) can be continuously divided into a combination of the smaller
parts of Fig. 3(i) or Fig. 3(ii), and an even smaller part of the form of
Fig. 3(iii), and Fig. 4. If this division is continued, one
can eventually break Fig. 3(iii) into a combination of several components of
Fig. 3(i) or Fig. 3(ii), Fig. 2(iv), and Fig. 4. As shown above,
all these contributions from Fig. 3(i), Fig. 3(ii), Fig. 2(iv), and Fig. 4
do not contribute to T-odd amplitudes. Therefore, the T-odd
amplitude from Fig. 3(iii) is zero.

Since an arbitrary possible T-odd forward scattering amplitude can be
obtained from combinations of Fig. 3(i) or Fig. 3(ii), Fig. 3(iii) and Fig. 4,
we conclude that
T-odd forward scattering amplitude of a polarized reaction
$1/2 + 1/2 \rightarrow 1/2 + 1/2$ within the MSMCKM is identically zero to all
orders. This implies that the T-odd total cross section of a polarized
reaction $1/2 + 1/2 \rightarrow 1/2 + 1/2$
is zero to all orders, by the optical theorem.

Because both T-odd/P-even and T-odd/P-odd amplitudes have the same CKM-matrix
factors, both T-odd/P-even and T-odd/P-odd amplitudes
should vanish to all order. We also note that
T-odd/P-even amplitude should be zero based on eq. 2, and the zero T-odd/P-odd
amplitude is due to the fact that the source of T-odd amplitude in the MSMCKM
is introduced by the phase in the CKM matrix.
The proof is therefore completed.

Two points need to be re-stated.

1) The above conclusion shows that a non-zero T-odd total cross
section of a polarized reaction $1/2 + 1/2 \rightarrow 1/2 + 1/2$
indicates the existence of additional T (or CP) violation source(s) besides
the phase in the CKM matrix. Furthermore, it is a null test in which a
high experimental accuracy can be achieved.

2) A zero T-odd total cross section does not indicate T (or CP)
conservation in the physical process, i.e.
a T (or CP) violation in a physical process is
a necessary but not sufficient condition for an existence of a T-odd
total cross section.

Ref. \cite{gxu} showed that the T-odd/P-odd total cross section of
a $1/2 + 1/2 \rightarrow 1/2 + 1/2$ reaction could be non-zero if the
Higgs sector contributes to T (or CP) violation. Thus, a measurement
of a T-odd/P-odd total cross section, proportionate to the
forward scattering amplitude, would indicate an
additional mechanism(s) of T (or CP) violation.
Based on the importance of T-odd total cross section measurement,
possibilities to carry out such measurements will be estimated in the
following.
These estimates are based on very limited knowledge of the Higgs
sector, so that the study is restricted to two Higgs doublets in which
T (or CP) symmetry is broken by neutral Higgs boson exchange. If there
are three Higgs doublets, both neutral and charged Higgs boson exchange
could break T (or CP) symmetry. This would enhance the signals obtained
from two Higgs doublets.

\section{Several T-Odd/P-Odd Processes}

The neutral scalar-quark interactive Lagrangian is given by,
\begin{eqnarray}
  {\mathcal L}_{\phi} & = &
      -{1 \over \sqrt{2} | \lambda_1|} \overline{D} m_D D \Phi_1
      +{i|\lambda_2| \over \sqrt{2} |\lambda_1|
      \sqrt{|\lambda_1|^2 +|\lambda_2|^2}} \overline{D} m_D \gamma^5 D \Phi_3
    \nonumber \\
    &   &
      -{1 \over \sqrt{2} | \lambda_2|} \overline{U} m_U U \Phi_2
      +{i|\lambda_1| \over \sqrt{2} |\lambda_2|
      \sqrt{|\lambda_1|^2 +|\lambda_2|^2}} \overline{U} m_U \gamma^5 U \Phi_3
    + h.c. \ .
\end{eqnarray}

At tree level, there could be three possible T-odd/P-odd forward
scattering amplitudes for $a(1/2) + b(1/2) \rightarrow a(1/2) + b(1/2)$,
shown in Fig. 5.

\begin{picture}(333,115)
  \multiput(5,100)(45,0){2}{\vector(1,0){22.5}}
  \multiput(27.5,100)(45,0){2}{\line(1,0){22.5}}
  \multiput(5,50)(45,0){2}{\vector(1,0){22.5}}
  \multiput(27.5,50)(45,0){2}{\line(1,0){22.5}}
  \put(58,65){$H$}
  \multiput(50,50)(0,10){5}{\line(0,1){5}}
  \put(50,100){\circle*{3}}
  \put(50,50){\circle*{3}}
  \put(50,71.5){\vector(0,1){3}}
  \put(2,105){$a$}
  \put(90,105){$a$}
  \put(2,35){$b$}
  \put(90,35){$b$}
  \put(45,22){(i)}
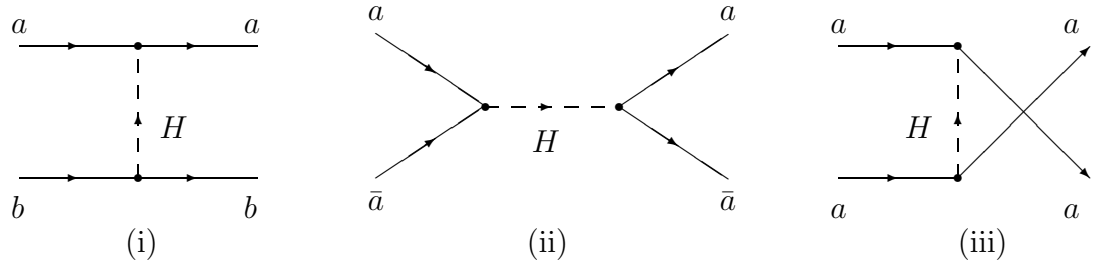
  \put(0,0){Fig. 5: The Higgs contributions to forward scattering amplitudes
  of $a+b\rightarrow a+b$}
  \put(140,105){\vector(3,-2){22.5}}
  \put(158.72,92.52){\line(3,-2){22.5}}
  \multiput(181.42,77)(10,0){5}{\line(1,0){5}}
  \put(140,50){\vector(3,2){22.5}}
  \put(158.72,62.48){\line(3,2){22.5}}
  \put(231.73,77){\vector(3,2){22.5}}
  \put(231.73,77){\vector(3,-2){22.5}}
  \put(250.71,89.5){\line(3,2){22.5}}
  \put(250.71,64.5){\line(3,-2){22.5}}
  \put(199,60){$H$}
  \put(181.42,77){\circle*{3}}
  \put(232,77){\circle*{3}}
  \put(203.5,77){\vector(1,0){3}}
  \put(137,110){$a$}
  \put(270,110){$a$}
  \put(137,38){$\bar{a}$}
  \put(270,38){$\bar{a}$}
  \put(197,22){(ii)}
  \put(315,100){\vector(1,0){22.5}}
  \put(337.5,100){\line(1,0){22.5}}
  \put(315,50){\vector(1,0){22.5}}
  \put(337.5,50){\line(1,0){22.5}}
  \multiput(360,50)(0,10){5}{\line(0,1){5}}
  \put(360,50){\vector(1,1){50}}
  \put(360,100){\vector(1,-1){50}}
  \put(340,65){$H$}
  \multiput(360,50)(0,10){5}{\line(0,1){5}}
  \put(360,100){\circle*{3}}
  \put(360,50){\circle*{3}}
  \put(360,71.5){\vector(0,1){3}}
  \put(312,105){$a$}
  \put(400,105){$a$}
  \put(312,35){$a$}
  \put(400,35){$a$}
  \put(360,22){(iii)}
\end{picture}

Here Fig. 5(i) is for the forward
scattering of two arbitrary spin-1/2 particles, Fig. 5(ii) is the
forward scattering for a spin-1/2 particle and its anti-particle, and
Fig. 5(iii) is for the forward scattering of two identical, spin-1/2 particles.
Based on the optical theorem, the T-odd/P-odd total cross sections of the
processes shown in Fig. 5 are given by,
\begin{eqnarray}
  \sigma_{t,5i}^{TP} & = & 0 \, , \nonumber \\
  \sigma_{t,5ii}^{TP} & = & {-1 \over v_{rel}}
    {m_a^2 |\lambda_i|v_j v_3 \over |\lambda_j|^2
    \sqrt{ |\lambda_1|^2 +|\lambda_2|^2}} {p_{az} \over p_{a0}}
    {m_H \Gamma \over (4p_{a0}^2 -  m_H^2)^2 +(m_H \Gamma)^2}
    s_{ax} s_{\bar{a}y} \, , \nonumber \\
  \sigma_{t,5iii}^{TP} & = &{-1 \over v_{rel}}
    {m_a^2 |\lambda_i|v_j v_3 \over |\lambda_j|^2
    \sqrt{ |\lambda_1|^2 +|\lambda_2|^2}} {p_{az} \over p_{a0}}
    { m_H \Gamma \over (a|\vec{p}_a|^2 + m_H^2)^2 +(m_H \Gamma)^2}
    s_{ax} s_{\bar{a}y} \, ,
\end{eqnarray}
where $v_{rel}$ is the relative velocity of two incoming particles,
the momenta are measured in CMS, and
$i = 1(2),\ j= 2(1)$ if the particle $a$ is in the $U(D)$ sector.
The effect of scalar exchange is assumed to be dominated
by the lightest neutral-scalar particle of mass $m_H$, i.e.,
\begin{eqnarray}
  <\Phi_i \Phi_j >_k \  \simeq {v_i v_j \over k^2 -m_H^2 +i\epsilon} \ .
\end{eqnarray}

Note from eq. (17) that $\sigma_{t,5ii}^{TP}$ reaches a
maximum at $p_{a0} = 0.5 m_H$, which corresponds to the resonance region of
the scattering, and that $\sigma_{t,5iii}^{TP}$ reaches a maximum
when $p_{a0}$
is slightly larger than $0$.\cite{expl3} Since there is no resonance in
Fig. 3(iii),
the maximum of $\sigma_{t,5iii}^{TP}$ can be orders of magnitude
smaller than the maximum of $\sigma_{t,5ii}^{TP}$.
Obviously, it is experimentally more
favorable to choose scattering channels and incoming particle momenta which
produce maximum T-odd/P-odd total cross sections. Further investigation
of the magnitude of the T-odd/P-odd total cross sections will require
knowledge of the Higgs sector and the masses of the quarks.
The following assumptions are adopted.

1) There is no prefered coupling of the Higgs to quarks
in the $U$ and $D$ sectors. This leads to,
\begin{eqnarray}
  |\lambda_1| \sim |\lambda_2| \sim (\sqrt{2} G_F)^{-1/2} = 246\, GeV\,.
\end{eqnarray}

2) The order of magnitude of $v_a v_b$ in eqs. (17)
and (18) is approximately 1.

3) The $u$ and $d$ quark masses are approximately $5\,MeV$.

Due to the small quark masses and large vacuum expectation values, the
couplings between quarks and the Higgs particle are very small, and small
T-odd/P-odd total cross sections are expected. Therefore, one should
attempt to find the largest open channels.
We consider several processes in the following.

\subsection{$p\bar{p}$ scattering}

We consider several factors in the T-odd/P-odd total cross section
for $p\bar{p}$ ~\cite{expl4}.

1) Since the valence quark composition of a proton is $uud$ and the valence
quark composition of an anti-proton is $\bar{u}\bar{u}\bar{d}$,
the dominant contributions to T-odd/P-odd total cross section
are from $\sigma_{t,5ii}^{TP}$ in eq. (17).

2) Valence quarks in a proton only contribute about $30\%$ of the proton
spin. Thus, it is assumed that each valence quark contributes about $10\%$
of the total spin.

3) Both valence and sea quarks in a proton contribute only about $50\%$ of
the proton total momentum. The most probable momentum for a valence quark in
a proton is about at $x = 0.15$.

Then the maximum total
cross section of $p\bar{p}$ is roughly given by,
\begin{eqnarray}
  \sigma^{TP}_{p\bar{p}} & \simeq &
    ({2\over 10}) \sigma^{TP}_{t,5ii} (u\bar{u}\rightarrow u\bar{u}) +
    ({1\over 10}) \sigma^{TP}_{t,5ii} (d\bar{d}\rightarrow d\bar{d})
    \nonumber \\
  & = & ({2\over 10}){-1 \over v_{rel}}
    {m_u^2 |\lambda_1|v_2 v_3 \over |\lambda_2|^2
    \sqrt{ |\lambda_1|^2 +|\lambda_2|^2}} {p_{uz} \over p_{u0}}
    {1 \over m_H \Gamma}
    s_{ux} s_{\bar{u}y} \nonumber \\
  &   & + ({1\over 10}) {-1 \over v_{rel}}
    {m_d^2 |\lambda_2|v_1 v_3 \over |\lambda_1|^2
    \sqrt{ |\lambda_1|^2 +|\lambda_2|^2}} {p_{dz} \over p_{d0}}
    {1 \over m_H \Gamma}
    s_{dx} s_{\bar{d}y} \, .
\end{eqnarray}

Using $m_H$ and $\Gamma$ given in ref. \cite{colphy}, and assuming
$\sigma_{t,p\bar{p}} \simeq 50\, mb$, an estimate of
$\sigma^{TP}_{p\bar{p}}$ and $A_{xy}$ is given in Table 1.
\begin{table}[htb!]
\caption{Maximum $\sigma^{TP}_{p\bar{p}}$ for various $m_H$} 
\begin{tabular}{|c|c|c|c|c|}
\hline
  $m_H\, (GeV)$     &  $\Gamma_H \,(GeV)$  &  $\sigma^{TP}_{p\bar{p}}\,(mb)$
    & $A_{xy}$ & beam energy in CMS ($GeV$)  \\ \hline
  $200$     &  $2 $  &  $8\times 10^{-14}$   & $2\times 10^{-15}$ &
    $667$ \\ \hline
  $400$     &  $25 $  &  $3\times 10^{-15}$   & $7\times 10^{-17}$ &
    $1333$ \\ \hline
  $600$     &  $100$  &  $6\times 10^{-16}$   & $1\times 10^{-17}$ &
    $2000$ \\ \hline
  $1000$     &  $450$   &  $8\times 10^{-17}$   & $2\times 10^{-18}$ &
    $3333$  \\ \hline
\end{tabular}
\label{table:table1}
\end{table}


One can see from Table 1 that the larger the Higgs
mass, the higher sensitivity is required to undertake a measurement.
In any event, it requires high accuracy and sensitivity
to measure such small cross sections.

Modern superconducting technology can measure current changes as low
as $10^{-8} \sim 10^{-9}\, A$. \cite{jc} If the lightest Higgs mass is not
too large, {\it e.g.} $m_H \sim 200\,GeV$, and the luminosities of the beams
are reasonably large, some of the small cross sections in Table 1 should be
measurable.

\subsection{$pp$ scattering}

  To estimate the T-odd/P-odd total cross section of $pp$,
a few factors will be considered.\cite{expl4}

1) The sea quarks in a proton are mainly found in the small $x$
region. Thus, the
major contributions to T-odd/P-odd forward scattering would most
likely occur in
valence quark collisions, {\it i.e.} $\sigma_{t,5iii}^{TP}$,
for beam energies below or around $m_H/2$. However,
if the beam energies were much beyond $m_H/2$, contributions from
sea quarks, {\it i.e.} $\sigma_{t,5ii}^{TP}$, could also be important. Only beam
energies below or around $m_H/2$ are considered.
For beam energies much beyond
$m_H/2$, one should refer to section 3.1.

2) The parton model is assumed to be valid at these beam energies.

3) As we only consider beam energies below or around $m_H/2$,
points 2) and 3)in section 3.1 remain valid.

The total cross section of $pp$ is roughly given by,
\begin{eqnarray}
  \sigma^{TP}_{pp} & \simeq &
    ({2\over 10}) \sigma^{TP}_{t,5iii} (uu\rightarrow uu) +
    ({1\over 10}) \sigma^{TP}_{t,5iii} (dd\rightarrow dd)
    \nonumber \\
  & \simeq & {-2 \over 10 v_{rel}}
    {m_u^2 |\lambda_1|v_2 v_3 \over |\lambda_2|^2
    \sqrt{ |\lambda_1|^2 +|\lambda_2|^2}} {p_{uz} \over p_{u0}}
    { m_H \Gamma \over (4|\vec{p}_u|^2 +m_H^2)^2 +(m_H \Gamma)^2}
    s_{ux} s_{\bar{u}y} + \nonumber \\
  &   & {-1 \over 10 v_{rel}}
    {m_d^2 |\lambda_2|v_1 v_3 \over |\lambda_1|^2
    \sqrt{ |\lambda_1|^2 +|\lambda_2|^2}} {p_{dz} \over p_{d0}}
    {m_H \Gamma \over (4|\vec{p}_d|^2 +m_H^2)^2 +(m_H \Gamma)^2}
    s_{dx} s_{\bar{d}y} \, .
\end{eqnarray}

Considering $m_H= 200\,GeV$ and $\Gamma=2\,GeV$\cite{colphy} and assuming
$\sigma_{pp} \simeq 50\,mb$, the estimated values of
$\sigma_{pp}^{TP}$ and $A_{xy}$ are given in Table 2.
\begin{table}[htb!]
\caption{\label{tab:table1}$\sigma^{TP}_{pp}$ vs beam energy in CMS for $m_H = 200\,GeV$}
\begin{tabular}{|c|c|c|}
\hline
    beam energy in CMS ($GeV$) & $\sigma^{TP}_{pp}\,(mb)$ & $A_{xy}$  \\ \hline
    $5$       & $8\times 10^{-18}$      & $2\times 10^{-19}$ \\ \hline
    $10$        & $8\times 10^{-18}$      & $2\times 10^{-19}$ \\ \hline
    $50$        & $5\times 10^{-18}$      & $1\times 10^{-19}$ \\ \hline
    $100$         & $2\times 10^{-18}$      & $4\times 10^{-20}$ \\ \hline
\end{tabular}
\end{table}


As one can see from the results in Table 2, the T-odd/P-odd total cross
sections for $pp$ are much smalller than for $p\bar{p}$.
This is understandable as there are no resonances in this channel.

One could also notices that the variations of $\sigma^{TP}_{pp}$ and
$A_{xy}$ versus beam energies are small. This is due to the large mass of
the Higgs particle, as compared to the incoming particle energies.

In general, both $\sigma^{TP}_{p\bar{p}}$ and $\sigma^{TP}_{pp}$ are
very small. However, as the above estimates are
based on a neutral scalar boson, other possible source(s) of
T (or CP) violation could be larger than these estimates.
A careful comparisons between $pp$ and $p\bar{p}$ T-odd/P-odd total cross
sections should provide more information of T (or CP) violation
mechanisms.

\subsection{$l\bar{l}$ and $ll$ scatterings}

If the coupling between the lepton sectors and the Higgs sector is similar
to the coupling between quark sectors and the Higgs sector, polarized
$l\bar{l}$ and $ll$ scattering can also have T-odd/P-odd total cross
sections. We consider the following points.

1) Leptons are elementary particles
and one does not need to consider the unpolarized and polarized
structure functions. Therefore, the result in eq. (17) can be directly used
for the $l\bar{l}$ and $ll$ T-odd/P-odd total cross sections.

2) The total cross sections for $l\bar{l}$ and $ll$ should be significantly
smaller than the $p\bar{p}$ and $pp$ cross sections since only electro-weak
interactions are involved. For beam energies which are
not in the $Z$ resonance region,
$\sigma_{l\bar{l}} \sim \sigma_{ll} \sim 10 \,\mu b$ is
assumed for simplicity\cite{expl5}.

The estimated $\sigma_{l\bar{l}}$($\sigma_{ll}$) and $A_{xy}$
are given in Tables 3-6.

\begin{table}[htb!]
\caption{Maximum $\sigma^{TP}_{e\bar{e}}$ for various $m_H$}
\begin{tabular}{|c|c|c|c|c|}
\hline
  $m_H\, (GeV)$     &  $\Gamma_H \,(GeV)$  &  $\sigma^{TP}_{e\bar{e}}\,(mb)$
    & $A_{xy}$ & beam energy in CMS ($GeV$)  \\ \hline
  $200$     &  $2 $  &  $3\times 10^{-15}$   & $6\times 10^{-13}$
    & $100$ \\ \hline
  $400$     &  $25 $  &  $1\times 10^{-16}$   & $2\times 10^{-14}$
    & $200$ \\ \hline
  $600$     &  $100$  &  $2\times 10^{-17}$   & $4\times 10^{-15}$
      & $300$ \\ \hline
  $1000$     &  $450$   &  $2\times 10^{-18}$   & $5\times 10^{-16}$
      & $500$  \\ \hline
\end{tabular}
\end{table}

\begin{table}[htb!]
\caption{$\sigma^{TP}_{ee}$ vs beam energy in CMS for $m_H = 200\,GeV$}
\begin{tabular}{|c|c|c|}
\hline
  beam energy in CMS ($GeV$)  & $\sigma^{TP}_{ee}\,(mb)$ & $A_{xy}$ \\ \hline
  $0.5$    & $3\times 10^{-19}$   & $6\times 10^{-17}$ \\ \hline
  $1$    & $3\times 10^{-19}$   & $6\times 10^{-17}$ \\ \hline
  $10$     & $3\times 10^{-19}$   & $6\times 10^{-17}$ \\ \hline
  $100$      & $7\times 10^{-20}$   & $1\times 10^{-17}$ \\ \hline
\end{tabular}
\end{table}

\begin{table}[htb!]
\caption{Maximum $\sigma_{\mu\bar{\mu}}^{TP}$ for various $m_H$}
\begin{tabular}{|c|c|c|c|c|}
\hline
  $m_H\, (GeV)$  &  $\Gamma_H \,(GeV)$  &  $\sigma^{TP}_{\mu\bar{\mu}}\,(mb)$
    & $A_{xy}$ & beam energy in CMS ($GeV$)  \\ \hline
  $200$     &  $\ \ 2$ & $1\times 10^{-10}$   & $2\times 10^{-8}$
    & $100$ \\ \hline
  $400$     &  $\ 25$  & $5\times 10^{-12}$   & $1\times 10^{-9}$
    & $200$ \\ \hline
  $600$     &  $100$   & $8\times 10^{-13}$   & $2\times 10^{-10}$
    & $300$ \\ \hline
  $1000$      &  $450$   & $1\times 10^{-13}$ & $2\times 10^{-11}$
    & $500$  \\ \hline
\end{tabular}
\end{table}

\begin{table}[htb!]
\caption{$\sigma_{\mu\mu}^{TP}$ vs beam energy in CMS for $m_H = 200\,GeV$}
\begin{tabular}{|c|c|c|}
\hline
  beam energy in CMS ($GeV$) & $\sigma^{TP}_{\mu\mu}\,(mb)$ & $A_{xy}$ \\ \hline
  $0.5$     & $1\times 10^{-14}$   & $2\times 10^{-12}$  \\ \hline
  $1$     & $1\times 10^{-14}$   & $2\times 10^{-12}$  \\ \hline
  $10$      & $1\times 10^{-14}$   & $2\times 10^{-12}$  \\ \hline
  $100$       & $3\times 10^{-15}$   & $6\times 10^{-13}$  \\ \hline
\end{tabular}
\end{table}

Similar to the results of baryon collisions, $l\bar{l}$ collisions have larger
T-odd/P-odd total cross sections if the Higgs sector is one of the T-violation
sources
and if the coupling between the lepton sectors and the Higgs sector
is similar to the one between the quark sectors and the Higgs sector.
Especially for the $\mu \bar{\mu}$ channel, its T-odd/P-odd total cross
section is
orders of magnitude larger than the corresponding $p\bar{p}$ channel
due to the larger masses of $\mu$ and $\bar{\mu}$.
If there were high energy polarized $\mu$ and $\bar{\mu}$ beams available,
the measurements of T-odd/P-odd total cross sections of
$\mu$ and $\bar{\mu}$
collisions should provide a sensitive test.

On the other hand, the T-odd/P-odd total cross section of $ll$ collisions is
an order(s)
of magnitude smaller than the corresponding $l\bar{l}$ total cross section.
These estimates are based on the assumption that the neutral Higgs particle
contributes to T (or CP) violation. If a small T-odd/P-odd cross
section of $ll$ collisions is measured, a careful comparison between $ll$
and the corresponding $l\bar{l}$ T-odd/P-odd total cross sections would
provide
additional information about T (or CP) violation.

\section{Summary and Future Prospect}

This note addresses the possibilities of searching for additional sources of
T (or CP)
violation through T-odd/P-odd total cross section
measurements. It shows the following.

\begin{enumerate}
\item A non-zero T-odd/P-odd total
cross section in the null test
$1/2 + 1/2 \rightarrow 1/2 + 1/2$ will indicate that there is(are)
additional source(s) of T (or CP) violation besides the phase in CKM matrix.
\item The contributions to T-odd/P-odd total cross
section from the Higgs sector can appear at tree level if the Higgs sector
contribute to T (or CP) violation, and the channels with resonance can be
measurable with modern technology if the lightest Higgs mass is not too large
and beam luminosities are reasonably large.
\item If the Higgs coupling to the
leptons is similar to the coupling to quarks, $A_{xy}$ in
both $l\bar{l}$ and $ll$ are larger than for $pp$ or $p\bar{p}$ due to
the smaller $l\bar{l}$ and $ll$ total cross sections, and $\mu{\bar{\mu}}$
provides the most sensitive channel due to the larger muon mass.
\item The present study only considers the coupling of neutral Higgs
particle as the additional source of T (or CP) violation. Actual
measurements could be larger if other mechnisms of T (or CP)
violation occur. A careful comparison between channels with and without
resonance could reveal mechanisms of T (or CP) violation beyond
current ideas.
\item Measurement of $A(x,y)$ in the total
cross section is
a null test and as such has the possibility to
give very accurate results.
\item The proposed measurements can provide
information on possible extentions of MSMCKM and the properties of vacuum.
\end{enumerate}

\noindent{\Large\bf Acknowledgments}
 
We would like to thank Professor Kwang Lau of Physics Department
at University of Houston for helpful discussions. We are specially
gratitide to Professor Ed V. Hungerford of Physics
Department at University of Houston for detailed review of the manuscript and
valuable contributions.


\begin{thebibliography}{}
\bibitem{gxu} Xu, Guanghua, Hungerford, Ed V., Nucl. Phys. B {\bf 649} (2003)
327-348.
\bibitem{weinbg} Weinberg, S. Phys. Rev.Lett. 19 1264-1266 (1967); Salam, A.
(1968). In Elementary particle physics (Nobel Symp No. 8). (ed. N. Svartholm).
Almqvist and Wilsell, Stockholm.
\bibitem{km} Kobayashi, M. and Maskawa, M., Prog. theor. Phys. {\bf 49} (1973)
652.
\bibitem{dolgov} Dolgov, A. D., Phys. Rep. {\bf 222} (1992) 309 and references
therein.
\bibitem{lee} Lee, T. D., Phys. Rep. {\bf 9C} (1974) 143.
\bibitem{weinberg} Weinberg, S., Phys. Rev. Lett. {\bf 37} (1976) 657.
\bibitem{weinberg2} Weinberg, S., Phys. Rev. Lett. {\bf 63} (1989) 2333;
Phys. Rev. D {\bf 42} (1990) 860.
\bibitem{conzett} Conzett, H. E., Phys. Rev. C {\bf 48} (1993) 423.
\bibitem{stapp} Stapp, Henry P., Phys. Rev. {\bf 103} (1956) 425;
Chou, Kuang-Chao, and Shirokov, M. I., Soviet Phys. JETP, {\bf 7} (1958) 851;
Csonka, P. L., Moravcsik, M. J., and Scadron, M. D., Ann. PHYS. {\bf 40} (1966)
100.
\bibitem{expl1} Here the $Z(\gamma)$ exchanges are excluded for they do
not introduce additional T-odd factors.
\bibitem{expl2} The only difference here from Fig. 3(i, ii) is that the
initial and final quarks
need to be summed over all components in the corresponding quark sector,
but this difference does not change the conclusions of eqs. (13) and (15).
\bibitem{expl3} The exact solution of $|\vec{p}_a|$ to have maximum
$\sigma_{5iii}^{TP}$ can be determined from this cubic equation
$64|\vec{p}_a|^3 +16(m_H^2+3m_a^2)|\vec{p}_a|^2+8m_a^2 m_H^2 |\vec{p}_a|+
m_a^2 m_H^2(m_H^2 +\Gamma^2)=0$.
\bibitem{expl4} This is just a rough estimate. A more accurate estimate
needs calculations based on up-to-date quark structure functions and polarized
quark structure functions. Sea quark contributions should also be
considered.
\bibitem{colphy} Vernon D. Barger, Roger J. N. Phillips, Collider Physics
(Addison-Wesley Publishing Company, 1987).
\bibitem{jc} J. Clarke, Phil. May. {\bf 13} (1986) 115;
\bibitem{expl5} For $l\bar{l}$ scattering, the maximum $\sigma_t^{TP}$ is at
the Higgs resonance region, but the total cross section is not expected to be
dominant in the Higgs resonance region due to the small coupling constants.
\end{thebibliography}
\end{document}